\documentclass{aa}                                

\usepackage{epsfig}     
\usepackage{graphicx,color}     
\usepackage{amssymb}            
\usepackage{url}                
\usepackage{amsmath}            
\usepackage{rotating}                   
\usepackage{float}                      
\usepackage{textcomp}
\usepackage{epstopdf}
\usepackage{dcolumn}
\usepackage{times}
\usepackage{tabularx}
\usepackage{hyperref}
\usepackage[normalem]{ulem}
\hypersetup{
    colorlinks,
    citecolor=blue,
    filecolor=blue,
    linkcolor=blue,
    urlcolor=cyan,
    menucolor=black}
\usepackage{soul} 
\usepackage[english]{babel}
\usepackage{booktabs}
\usepackage{xspace}
\usepackage{gensymb}
\usepackage{orcidlink}

\newcommand{\vx}{${v_x}$\xspace}
\newcommand{\vy}{${v_y}$\xspace}
\newcommand{\vz}{${v_z}$\xspace}

\begin{document}

\title{Polarization of decayless kink oscillations in a 3D MHD coronal loop model}

\author{
Sudip~Mandal\inst{1}\orcidlink{0000-0002-7762-5629},
Cosima~Breu\inst{2}\orcidlink{0000-0003-2105-238X}
\and
Hardi~Peter\inst{1,3}\orcidlink{0000-0001-9921-0937}}
\institute{
Max Planck Institute for Solar System Research, Justus-von-Liebig-Weg 3, 37077, G{\"o}ttingen, Germany \\
\email{smandal.solar@gmail.com}
\and
The University of Graz, Universit{\"a}tspl. 3, 8010 Graz, Austria
\and
Institut f{\"u}r Sonnenphysik (KIS), Georges-Köhler-Allee 401a,
79110 Freiburg, Germany
}

\abstract
{Decayless kink oscillations are frequently observed in solar coronal loops and are considered potential contributors to coronal heating. Despite the ubiquity of this wave phenomenon, its driving mechanism remains unclear. Studies to derive the polarization state of these oscillations, which would be a key to identifying the drivers, have been limited due to observational constraints. We analyze a 3D MHD simulation of coronal loops using the MURaM code. Synthetic extreme ultraviolet (EUV) emission maps, combined with velocity diagnostics, are used to identify and characterize transverse wave motions in the simulated loop structures. This is the first demonstration of decayless kink waves emerging self-consistently in a 3D MHD loop-in-a-box model. The simulation produces persistent, low-amplitude, decayless kink oscillations that closely match observed properties. These oscillations arise spontaneously, without any imposed periodic driver, and exhibit clear linear polarization with oscillation planes not aligned to the principal axes. The observed coherency of linear polarization with oscillation cycles favors a self-sustained or quasi-steady type wave driver over a stochastic or broadband source.}

   \titlerunning{Decayless kink oscillations in coronal loops without imposed driving}
   \authorrunning{Sudip Mandal et al.}
   \maketitle
 
\section{Introduction} \label{sec:intro}
Kink oscillations in coronal loops were first reported by \cite{1999Sci...285..862N,1999ApJ...520..880A}, where loops were observed swinging side to side in response to flare-like perturbations occurring close to the loops. A characteristic of these oscillations is that they damp rapidly, typically within a couple of wave periods. This is in contrast to the small-amplitude kink waves discovered more recently (e.g., \citealp{2012ApJ...759..144T,2013A&A...560A.107A}), which maintain a nearly constant amplitude over multiple cycles, which is why they have been termed ``decayless'' oscillations, in contrast to the earlier type known as ``decaying'' oscillations. More importantly, these decayless waves exist without the presence of any obvious transient driver and identifying their driver(s) from observations has remained a challenge thus far \citep{2022A&A...666L...2M}. One indirect method to investigate the wave drivers is to examine the polarization properties of these decayless kink waves. For instance, if these waves are driven by a coherent driver (e.g., background flows; \citealp{2020ApJ...897L..35K}), the resulting wave polarization would be strictly linear at all time. In contrast, if the wave driver operates randomly (e.g., random footpoint driving; \citealp{2020A&A...633L...8A}), the resulting wave polarization would also be linear, but the orientation of the polarization would vary stochastically over time. From an observational standpoint, determining the polarization state of decayless waves accurately is challenging. It often necessitates the use of multiple type of instruments (imaging and spectroscopic) or stereoscopic measurements of a loop, ensuring that the alignment minimizes projection effects. As a result, such reports are rare in the literature. Recently, \cite{2023NatCo..14.5298Z} demonstrated through stereoscopic observations that the polarization of these waves is most likely linear in nature and that this polarization state is maintained over multiple oscillation cycles.

In recent years, significant efforts have been made to investigate the properties of decayless waves using 3D magnetohydrodynamic (MHD) simulations. However, most of these studies rely on an explicitly prescribed wave driver to generate these oscillations. For example: (i) a steady background flow across the loop \citep{2020ApJ...897L..35K,2021ApJ...908L...7K}; (ii) a harmonic driver near the loop footpoints that imposes a prescribed period \citep{2023ApJ...949...38S,2023ApJ...955...73G}; or (iii) random foot-point driving using a verity of driver profiles \citep{2019FrASS...6...38K,2021ApJ...908..233S,2024A&A...681L...6K}. In contrast, \citet{2021A&A...647A..81K} recently presented a self-consistent 3D MHD model that provides evidence of both decaying and decayless kink oscillations along selected field lines. Their study, however, neither identifies loop structures as a whole (ensemble statistics) nor explores wave polarization. Additionally, as the simulation box in their simulation is relatively small, the field lines examined measure approximately only 15 Mm in length, making them representative of short, low-lying loops. In this paper, we present evidence of decayless kink waves in longer coronal loops (approximately 50 Mm in length), which emerge naturally from a 3D MHD simulation. Additionally, we investigate the polarization of these waves because it can serve as a diagnostic tool for identifying the wave driver.

\begin{figure*}[!ht]
\centering
\includegraphics[width=0.80\textwidth,clip,trim=0cm 0cm 0cm 0cm]{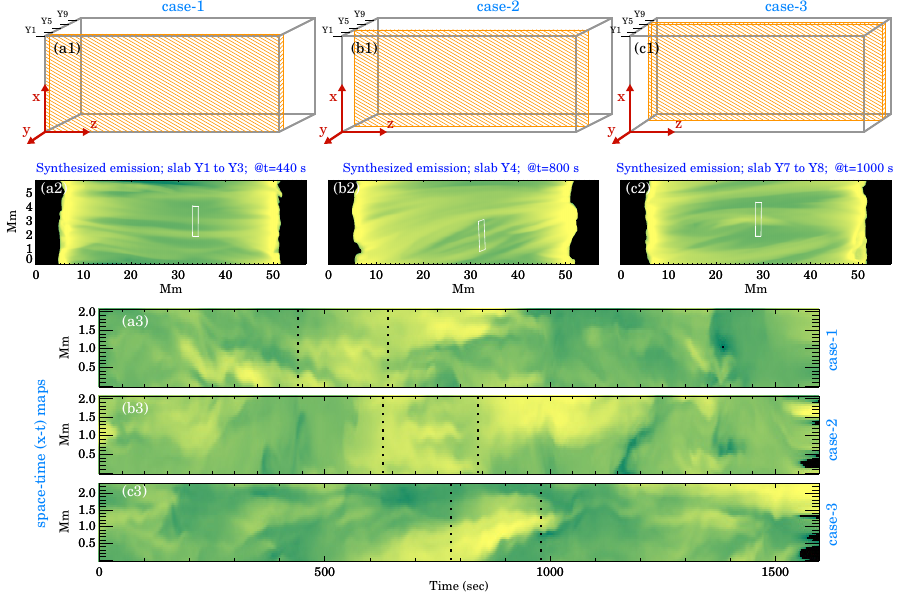}
 \caption{An overview of the $x$-$t$ maps generated from the synthetic 171~{\AA} image sequence. Panel a1 illustrates the schematic of the simulation domain, where the orange hatched area outlines the volume used to generate the synthetic image of case 1, shown in panel a2. The white rectangular box in panel a2 indicates the location and extent of the artificial slit used to create the space-time ($x$-$t$) map displayed in panel a3. The two vertical lines in this $x$-$t$ map define the time window during which decayless oscillations are observed . Cases 2 and 3 are presented in panels b1 to b3 and c1 to c3, respectively. }
\label{fig:context_plot}
\end{figure*}

\section{Numerical model}\label{sec:model}
In this study, we utilize the output from the coronal loop model presented by \cite{2022A&A...658A..45B}. This model employs the 3D radiative MHD MURaM code \citep{2005A&A...429..335V} that includes an extension to the corona \citep{2017ApJ...834...10R}. In this simulation, a coronal loop is modeled as a straight flux tube, with both ends anchored in the photosphere. The atmosphere between these ends includes the chromosphere and the corona. In the setup, loop curvature is neglected, as it is expected to introduce only minor corrections to the global kink-mode properties considered here for loops with large aspect ratios \citep{2004A&A...424.1065V}, while gravitational stratification is explicitly included (Eqn. 6 of \citealp{2022A&A...658A..45B}). 
The simulation is driven by a selfconsistent near-surface magnetoconvection at the loop footpoints. For comprehensive details about the model setup, including boundary conditions and physical parameters, we refer to \cite{2022A&A...658A..45B}. Here, we highlight the critical parameters relevant for understanding the results presented in this study. The simulation domain has dimensions of 6$\times$6$\times$57~Mm, with a grid spacing of 60~km in each direction (which corresponds to 100$\times$100$\times$950 grid points). Excluding the convection zone located at the bottom of each footpoint, the effective loop length is approximately 50~Mm. The average unsigned surface magnetic field is around 70 G, which is typical of weak plage regions on the Sun. After the model reached some state of equilibrium (still with considerable dynamics), in the corona the average temperature is in the range of 0.9-4 MK while electron densities lie between 1$\times$10$^{8}$~cm$^{-3}$ and 5$\times$10$^{8}$~cm$^{-3}$. The simulated data set has a write-out cadence of the snapshots of 2~s over a total duration of 27 min.

To asses the  visibility of the loops in images captured in typical extreme UV (EUV) image data, we synthesize the EUV emission that would be expected from the model. Here we concentrate on the 171\,{\AA} channel as observed by the Atmospheric Imaging Assembly \citep{2012SoPh..275...17L}. This captures mainly plasma around 1 MK and is similar to the passband used in the High Resolution Telescope of the Extreme UV Imager on Solar Orbiter  \citep{2020A&A...642A...8R}). The synthesized emission facilitates a straightforward comparison with actual observations. Further details about the synthesis procedure is available in \cite{2022A&A...658A..45B}.

\section{Method}\label{sec:method}
We base our analysis on the appearance of loops observed in the synthetic 171~{\AA} image sequence. The aim is to identify the volume that contributes to the observed emission patterns of the oscillation. To achieve this, we follow the traditional method of tracking transverse oscillations in synthetic coronal images. This involves placing an artificial slit perpendicular to the length of the oscillating loop to generate a space-time ($x$-$t$) map \citep{2021A&A...652L...3M}.

However, applying this method to the image sequence which is created by integrating the emission over the entire y-direction (the line-of-sight direction) of the simulation box, introduces inherent uncertainty in accurately locating the oscillating structures in 3D. Like on the real Sun this uncertainty arises from the optically thin nature of coronal emission lines, such as present in the 171~{\AA} band used in this study.

To effectively identify and isolate an oscillating loop, we divide the y-direction (line-of-sight) into slabs, each 600 km (10 pixels) deep, and compute synthetic emission for each of these y-slabs. We then manually check which y-slabs contribute to the emission of the loop of interest. For example, the $x$-$t$ map shown in Fig. \ref{fig:context_plot}a3 was created by placing an artificial slit over a specific loop in the emission map illustrated in Fig.~\ref{fig:context_plot}a2. The emission map itself was generated by integrating the pixels from y=0 to y=29, corresponding to slab-y1 to slab-y3 in the simulated domain (Fig.~\ref{fig:context_plot}a1). We observed that the oscillations in the $x$-$t$ map, as well as the loop in question, disappear when we consider only the emission from slab-y4 and beyond. This approach allows us to isolate the volume contributing to the observed emission. There are several oscillating loops in the simulation; however, we have chosen only four of them for demonstration purposes in this study (Fig. \ref{fig:context_plot}). In the following sections, we furthermore focus on selected time windows and investigate these oscillations in detail.

\section{Results} 
\subsection{Decayless signatures}\label{sec:decayless}

\begin{figure}[!ht]
\centering
\includegraphics[width=0.49\textwidth,clip,trim=0cm 0cm 0cm 0cm]{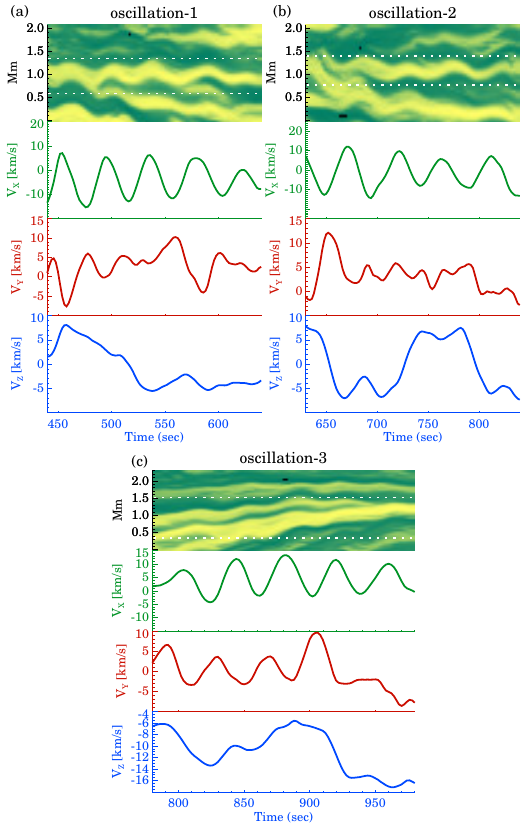}
\caption{Signatures of decayless oscillations in emission $x$-$t$ maps and in the velocity components. Column-a presents signatures of oscillation-1 (from case-1). In this column, the top row displays a zoomed-in $x$-$t$ map from Fig.~\ref{fig:context_plot}, while the next three rows illustrate the time evolution of the velocity components \vx, \vy and \vz, respectively. The white horizontal lines indicate the $\delta$x extent of the 3D volume from which the velocity components were obtained. See Sect.~\ref{sec:decayless} for more details. The other two columns (b and c) present information for oscillations 2 and 3 in the same format.}
\label{fig:param_plot}
\end{figure}

To analyze the decayless wave patterns we zoom into the regions of the space-time maps that cover only a few instances of decayless waves in neighboring threads. Examples are shown in Fig.~\ref{fig:param_plot}, where we present zooms into the three $x$-$t$ maps from Fig.~\ref{fig:context_plot}. To enhance the visibility of the oscillating structures, we applied the Multi-scale Gaussian Normalization (MGN) filter \citep{2014SoPh..289.2945M} when producing these $x$-$t$ maps. In all three maps, we identify several threads exhibiting persistent oscillations. We emphasize that these decayless oscillations appear self-consistently from the model, without any prescribed driver enforced.

Our analysis of these oscillating threads (see Appendix~\ref{app:detrending}) indicates that their periods range from 40 to 60 seconds, with typical amplitudes around 0.1 Mm. The periods we observed in this study are consistent with those reported in earlier work. For example, applying the relationship between period and loop length for standing modes of decayless oscillation in a loop measuring 50 Mm, we expect the period in our synthetic observations to be approximately 55 seconds \citep{2024A&A...685A..36S}. This closely aligns with the values we measure. Furthermore, the wave amplitude of 0.1~Mm is also typical for loop observations \cite[e.g.][]{2015A&A...583A.136A}. Lastly, by examining oscillation patterns derived from slits positioned at different locations along the loop, we confirm that these waves are the fundamental mode of standing, decayless oscillations (see Appendix~\ref{app:standing}).

The question that remains is to determine what actually drives these decayless oscillations. As mentioned in the introduction, tracking field lines in both space and time is challenging in these simulations. Therefore, we seek an indirect but effective proxy to better understand the characteristics of the driver. One such proxy is wave polarization and to do that, we require the time evolution of the 3D velocity components associated with the oscillating plasma in the loop. From the previous analysis (mentioned in Section~\ref{sec:method}), we obtained $\delta y$ values, while the location of the slits provided $\delta z$ values. To determine the $\delta x$ values, we manually identify the spatial extent of oscillations in the $x$-$t$ maps by plotting horizontal lines across each $x$-$t$ map that fully cover the oscillating loop during the selected period, while ensuring minimal overlap with other structures (see Fig.~\ref{fig:param_plot})\footnote{Our results are not sensitive to the exact values of $\delta x$, $\delta y$, and $\delta z$ used to derive the average velocities \vx, \vy, and \vz that we analyze here.} Through this exercise, we identify the 3D volume that contains the section of the loop mapped into a $x$-$t$ map. The velocities we quote in the following sections are all averaged quantities within this identified volume.

In all three cases, we observe clear oscillatory signals in \vx that do not exhibit any substantial decay over time for multiple cycles. While some degree of decay or modulation is present, it is not close to the exponential decay typically observed in decaying oscillations. The presence of such oscillations in \vx aligns well with our findings in the synthesized emission maps. Furthermore, periodicities are found in the FFT power spectrum of \vx, which shows clear a peak around 40~seconds, a  value quite similar to the 43~seconds periodicity we obtained from the $x$-$t$ map (see Appendix~\ref{app:detrending}). 

It is not surprising that the oscillations we show here in Fig.~\ref{fig:param_plot} is most pronounced in the $x$ direction and therefore in \vx. This is simply because we conduct this analysis for instances of decayless waves that appear prominently in planes where the line-of-sight is perpendicular to the $x$ direction. What is more surprising is that the oscillatory nature of the \vy and \vz curves is evident in all cases. This, combined with the oscillation in \vx, could indicate two possibilities: 1) the observed kink oscillations are not linearly polarized, i.e., they may be elliptically or circularly polarized; or 2) they are inherently linearly polarized, but the loop oscillation plane is tilted with respect to the line of sight, resulting in oscillatory signals in the other velocity components. To investigate which of the two possibilities applies we study the hodograms for these waves.

\subsection{Hodograms}\label{sec:hodograms}
One effective way to visualize the relationships between different velocity components is through hodograms, which are parametric plots that illustrate the trajectory of a point with respect to two or more variables. To facilitate this, we first detrend the \vx, \vy, and \vz curves so that the center of the hodograms does not show any artificial shifts over time. The detrended curves are presented in Appendix~\ref{app:detrending}. 
In Fig.~\ref{fig:hodograms} we show the hodograms for each pair of \vx, \vy, and \vz for all three oscillations displayed in Fig.~\ref{fig:param_plot}. 

The \vx vs. \vy hodograms in each oscillation exhibit loopy elliptical structures that remain mostly confined within a narrow band. In the case of \textit{oscillation-3}, we notice that the ellipses rotate slightly over time. The \vx vs. \vz pairs also display elliptical trajectories, but they are much narrower in width, as the \vz values are significantly smaller than the \vx values. A similar pattern is observed in the \vy vs. \vz hodograms.

\begin{figure}[!ht]
\centering
\includegraphics[width=0.49\textwidth,clip,trim=0cm 0cm 0cm 0cm]{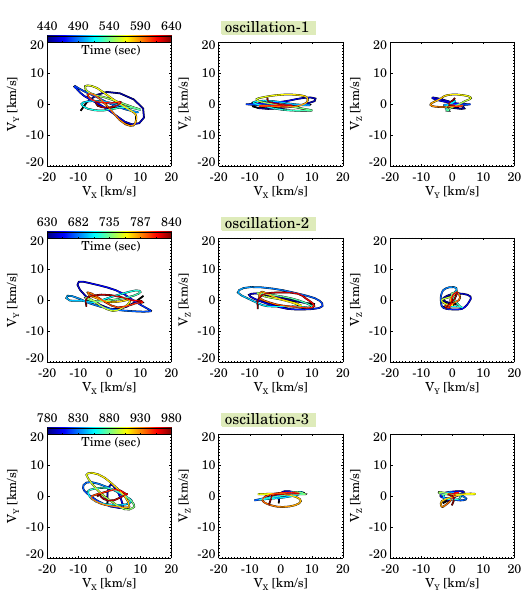}
\caption{Hodograms of the three decayless kink oscillations. In each instance, the colors in these hodograms represent the flow of time. An animated version is available \href{https://drive.google.com/drive/folders/1ul9ZDnsz77p9y-cPO5IVPfcqABkThlZG?usp=sharing}{online}.}
\label{fig:hodograms}
\end{figure}

Combining the information from all three hodograms for each oscillation, we find that all orbits in these plots are closed, which is characteristic of motion within a single plane. For intrinsically circularly polarized kink waves, we would expect to see elliptical (or circular) patterns that share nearly identical properties, such as ellipticity or time evolution, across all three hodograms. The variation in ellipticity observed in the 2D projections aligns most consistently with linearly polarized kink waves oscillating in planes that are tilted relative to all three axes. Hence we conclude that the decayless oscillations we see here are linearly polarized.

\section{Discussion}\label{sec:discussion}
In this study we set out to explore the existence of decayless kink oscillations in loops in a 3D MHD simulation. We presented multiple examples of such oscillations that arise self-consistently, without any imposed driver in the simulation setup. All those decayless cases are fundamental mode standing waves, with the antinode located at the loop apex and the nodes positioned at the footpoint.  Let us examine few of our principal findings:

 A fundamental characteristic of these waves is their ubiquity. In nearly all instances, we observed the wave to persist for approximately five cycles. This aligns well with the results of most observational investigations \citep{2013A&A...560A.107A,2015A&A...583A.136A,2022A&A...666L...2M,2024A&A...685A..36S}, although few reports mention decayless oscillations exceeding 30 cycles \cite[e.g.,][]{2022MNRAS.513.1834Z}. Nonetheless, whenever these oscillations are detected for fewer than five cycles, it has been proposed that the physical state of the oscillating plasma (including its temperature or density) has altered, making the oscillating loop undetectable by the specific instrument filter or line employed for observation. Therefore, in principle, the loop may still be oscillating, but we cannot perceive it. This hypothesis, however, does not match our cases, as we observed that the velocity rapidly diminishes after exhibiting four or five cycles of decayless signatures. This indicates that the field lines themselves are no longer oscillating. A potential explanation is that during these oscillations, the loop has gradually deviated from the particular 3D volume from which we are extracting the velocity component (we observe a slight trend in \vy in all three instances), and therefore, the abrupt disappearance may well be a manifestation of that behavior.\\

Another important diagnostic of decayless kink oscillations is the velocity amplitude, which constrains the wave energy content and is relevant to assessing their potential contribution to coronal heating when combined with an estimate of the effective dissipation rate.  In our simulation, the \vx values (representing the plane of sky motions) typically range around 10 km/s, while the line-of-sight (LOS) component \vy varies between 5 to 10~km~s$^{-1}$. These values are broadly consistent with those typically inferred from imaging and spectroscopic observations of decayless kink oscillations \citep[e.g.,][]{2012ApJ...759..144T,2013A&A...560A.107A}. Overall, our simulation framework successfully reproduces transverse motions of the same order as observationally found.\\ 
    
 Lastly, we discuss about identifying the driver of the oscillations observed in the simulation. One challenge in identifying the driver is that it can operate at any point along the length of a magnetic field line, from the photosphere to the corona (i.e., from one footpoint to the other), and still produce decayless oscillations \citep{2021SSRv..217...73N}. This makes the task quite complex, even in a simulation. As seen in the animation, the loops in this simulation are highly dynamic, drifting while oscillating, and presumably reconnecting and changing their connectivity. Consequently, tracking field lines throughout their entirety and throughout time presents significant challenges in identifying the driver. On the other hand, studying the properties of wave polarization offers an indirect method to infer characteristics of the driver without the need to track a loop along its entire length. All individual cases we studied here indicate that these decayless kink waves are largely linearly polarized, with their oscillation planes tilted relative to the principal axes. The presence of linearly polarized waves suggests that the driver preferentially excites waves in a specific direction. One potential source could be the self-oscillation model proposed by \citep{2016A&A...591L...5N}. In this model, atmospheric flows (e.g., along supergranular boundaries) with a much longer characteristic timescale than the oscillation period drive these waves, a conclusion also put forward by \cite{2023NatCo..14.5298Z}. Although in case of \textit{oscillation-2} and \textit{oscillation-3}, there appears to be a marginal rotation of the oscillation planes with each oscillation cycle, the patterns are still organized and stable. These features point to the fact that a truly random or broadband footpoint motion is not a primary driver in these loops as such motions would yield disordered, time-varying hodograms with inconsistent shape and direction. It is important to note that our simulation setup, by design, lacks large-scale supergranular flows, as the lower boundaries are only 6 × 6 Mm in size. Therefore, there must be flows at other spatial scales that contribute to the decay-less oscillation observed in our simulation. Finally, the lack of circularly polarized waves indicates that twisting or swirling motions are probably not responsible for these oscillations.  This is particularly intriguing given a comparable 3D simulation presented in \cite{2023A&A...675A..94B} reveals the existence of swirls in the lower atmosphere that pump a substantial amount of Poynting flux into the corona. Consequently, it remains unclear why these decayless oscillations fail to have any significant influence of such swirling motions. On the other hand, the elliptical shapes we observed may indeed signify that. Further investigations are necessary to examine this facet.

 Another notable aspect is that the oscillation planes in each case are tilted, which may indicate the presence of twist or writhe in the magnetic field. This twist could cause the loop to oscillate preferentially along a direction that is not perpendicular to the loop plane \citep{2012A&A...548A.112T}. In fact, these authors suggest that a varying twist in the field could also produce a variation in polarization, a phenomenon we observed in some of the hodograms in Fig~\ref{fig:hodograms}. In this context, we note that \cite{2022A&A...658A..45B} mentioned the presence of twist in several loop structures within this simulation setup (see Fig. 2 of their paper), which could indeed support our previous argument. The slight but consistent tilt of the \vx-\vy hodograms for each oscillation may also have the influence of the straightened loop geometry chosen for this simulation as well as the loop density contrast \citep{2006ApJ...650L..91T}. 
Finally, we note that the primary diagnostics for decayless kinks discussed above, namely period, phase, and polarization, are influenced by loop-scale dynamics and are well resolved at our grid spacing of 60 km (consisting of hundreds of cells along a loop of approximately 50 Mm). However, in the MURaM simulations, numerical diffusion affects structures smaller than about five grid cells (approximately 300 km; \citealp{2025MNRAS.537.2835B}). Therefore, while we consider our results to be robust regarding the overall kink behavior, we do not claim convergence for the small-scale dissipation or heating.

 \section{Prospects}\label{sec:limitation}

Although polarization states of the oscillation can reveal whether the wave driver is random or steady, they are unable to independently reveal some of the most important details about the driver. For example, the spatial placement of the driver—that is, whether it is acting along the length of a loop or close to its footpoints—is not uniquely determined by polarization states.  Additionally, we require further investigations on the following facets of our findings:  (i) Is the abrupt lack of oscillations we noticed truly a result of the loop exiting the analysis box we chose?  (ii) Does twisting inside the loops cause the oscillation polarization to be similar? (iii) Do the wave properties change if we would alter the photospheric magnetic environment of the numerical model (e.g., from plage to sunspot)? 

The core approach in addressing the above questions requires tracking the field lines associated with the oscillating loops. Our current analysis is based on the visibility of loops in synthetic images analogous to observations, which comes with certain limitations. For example, due to a number of changes in the plasma parameters, a loop may move in or out of the visibility range of a specific filter's bandpass. Therefore, our current method only allows us to capture a part of the loop's evolution, rather than the whole process. Additionally, in our current approach, we are extracting information from a volume that is fixed in time. However, in reality, a loop may move laterally over time, meaning it may not always be part of that volume (even though we took care when selecting the volume). Additionally, due to the dynamic appearance of these loops in the model, reliable field-line tracking becomes challenging as we approach the loop footpoints. This is particularly important for several wave properties. For instance, the presence of harmonics can only be meaningfully established by comparing oscillations at the loop apex with those in legs of the loop \citep{2018ApJ...854L...5D}. In short, the field lines covering each oscillating structure must be followed in space at every time step. This task is non-trivial, as field lines often reconnect along their lengths, making them harder to follow over time \citep{2021A&A...647A..81K}. Nonetheless, a study following the existing field-line tracing codes (such as one the one used in \citealp{2022A&A...661A..94C}) is currently underway.

\section{Conclusion}
This study provides evidence for the existence of decayless kink waves in long coronal loops. These waves emerge naturally from a 3D magnetohydrodynamic simulation. We further established the linear polarization states of these oscillations and emphasized that they favor a self-sustained or quasi-steady wave driver rather than a stochastic or broadband source.

\begin{acknowledgements}
We thank the anonymous reviewer for the encouraging comments and helpful suggestions. The work of S.M. was funded by the Federal Ministry for Economic Affairs and Climate Action (BMWK) through the German Space Agency at DLR based on a decision of the German Bundestag (Funding code: 50OU2201).
\end{acknowledgements}
\bibliography{kink_muram}
\bibliographystyle{aa}

\begin{appendix}
\onecolumn
\section{An intriguing example}\label{app:decaying}
\begin{figure*}[!ht]
\centering
\includegraphics[width=0.90\textwidth,clip,trim=0cm 0cm 0cm 0cm]{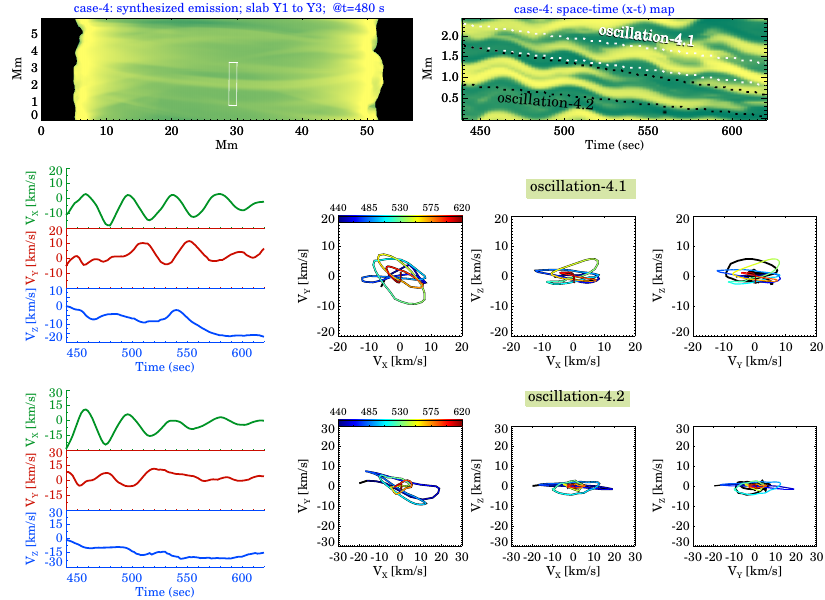}
\caption{An example of co-existing decaying and decayless kink oscillations in neighboring loops. Top row shows the synthetic 171~{\AA} image with the white rectangular box outlining the position of the artificial slit used to generate the $x$-$t$ map shown in next panel. The slanted white and black dotted lines mark the $\delta$x values for the two selected oscillations. For more information on this, see Sect~\ref{sec:decayless}. The middle row displays the \vx, \vy and \vz curves and the hodograms constructed for $\textit{oscillation-4.1}$. The same but for $\textit{oscillation-4.2}$ is shown in the bottom row panels. }
\label{fig:decaying}
\end{figure*}

Not all the transverse oscillations observed in the simulation are decayless. Among them, we present an intriguing case in Fig.~\ref{fig:decaying}. In this figure, oscillations, marked as \textit{oscillation-4.1} and \textit{oscillation-4.2}, are seen in two adjacent loops. However, their appearance in the $x$-$t$ map reveals that \textit{oscillation-4.1} is decayless in nature, while \textit{oscillation-4.2} exhibits signs of decay. This distinction becomes increasingly evident when we analyze the velocity components; we find that the velocity component \vx of \textit{oscillation-4.1} shows only a slight decrease over time, whereas \vx of \textit{oscillation-4.2} displays a rapid decay, resembling typical decaying kink oscillations. Constructed hodograms appear similar, further suggesting that both these these oscillations are most likely linearly polarized. Therefore, we have identified a case in which transverse oscillations occur simultaneously in two adjacent loops yet exhibit drastically different decay properties. Given that these two loops are most likely positioned at neighboring sites at the photospheric level, it is reasonable to conclude that they likely experience comparable footpoint driving conditions.  It is indeed puzzling that they display distinct signatures of entirely contrasting regimes of kink oscillation and that too, simultaneously.

\section{Detrended curves and periodicities}\label{app:detrending}
\begin{figure}[!ht]
\centering
\includegraphics[width=0.98\textwidth,clip,trim=0cm 0cm 0cm 0cm]{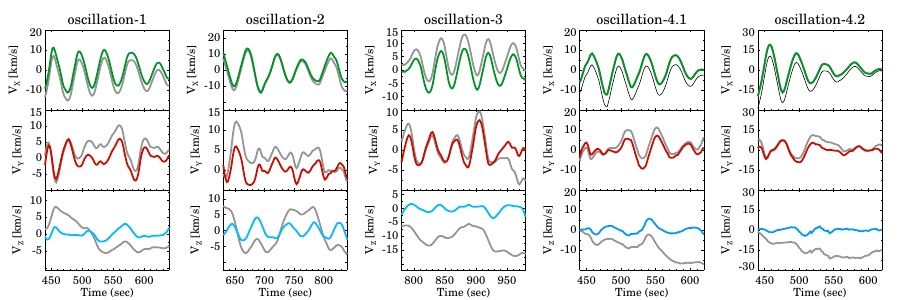}
 \caption{Original and detrended velocity curves for each oscillation. For a given oscillation, the original curves are shown in grey, while the detrended \vx, \vy, and \vz curves are displayed with green, red, and blue lines, respectively.}
\label{fig:detreded}
\end{figure}

As mentioned in Section~\ref{sec:hodograms}, the hodograms presented in Fig.~\ref{fig:hodograms} are constructed using the detrended velocity curves \vx, \vy, and \vz. This same approach was also utilized for the hodograms shown in Fig.~\ref{fig:decaying}. To determine the trend of a curve, we fit it with a best-fit polynomial (ranging from degree 1 to 5) and then subtract this fitted polynomial to obtain the final detrended curve. In Fig.~\ref{fig:detreded}, we provide all the velocity curves for the oscillations discussed in this study. 

\begin{figure}[!ht]
\centering
\includegraphics[width=0.80\textwidth,clip,trim=0cm 0cm 0cm 0cm]{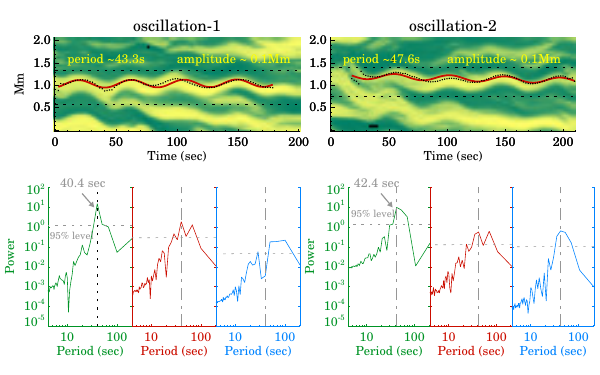}
 \caption{Estimation of oscillation periods based on the $x$-$t$ maps and the velocity components. The top row of panels displays the $x$-$t$ maps for $\textit{oscillation-1}$ and $\textit{oscillation-2}$. In these maps, dotted circles indicate the locations of the centers of the oscillating threads, while solid red curves represent the best-fit function (\ref{equation1}) to those points. The estimated periods and amplitudes are shown on the panels. The bottom row of panels presents the FFT power spectrum for each velocity component: \vx in green, \vy in red, and \vz in blue. The vertical dotted lines in these panels mark the locations of the dominant periods of \vx, while the horizontal dashed lines outline the 95\% significance level.}
\label{fig:fft}
\end{figure}

The next step was to compute the periodicities present in these detrended velocity curves. We did this by performing a Fast Fourier Transform (FFT) on each curve. The results are presented in the bottom row of Fig.~\ref{fig:fft} for two cases. The peak in \vx (the plane-of-sky component) shows the strongest and most distinct feature in the FFT power spectrum, which is expected. Peaks in \vy at the same period as \vx are also observed, albeit much weaker. This pattern is similarly seen in \vz. To compare the oscillations observed in the velocity components (especially \vx) with those in the emission $x$-$t$ maps, we fit the oscillating threads in the $x$-$t$ maps using the following approach: first, we identify the location of an individual strand in the $x$-$t$ map at each time step by fitting a Gaussian along the transverse direction of that strand. We then fit these derived positions of the strand as a function of time using the equation:
\begin{equation}
\label{equation1}
\centering
 y(t)=A \left (\sin{\dfrac{2\pi t}{P}+\phi} \right)+c_1 t+c_0,
\end{equation}
where $A$ is the oscillation amplitude, $P$ represents the period, $\phi$ is the phase, and  $c_{0}$ and $c_{1}$ are constants. Results based on these fits are shown in the top row of Fig.~\ref{fig:fft}.

\section{Standing wave signatures}\label{app:standing}
\begin{figure}[!ht]
\centering
\includegraphics[width=0.80\textwidth,clip,trim=0cm 0cm 0cm 0cm]{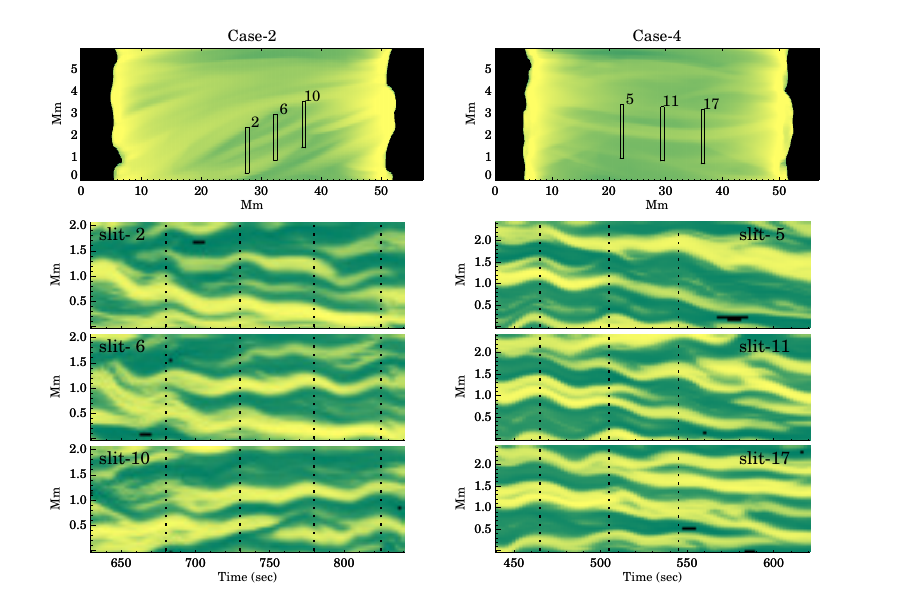}
 \caption{Identification of the wave mode through $x$-$t$ maps. The left column shows the emission map (top panel) followed by $x$-$t$ maps derived from the three slits indicated in the white rectangular boxes on the emission map. The dotted vertical lines mark the locations of the wave peaks and serve as a visual guide to identify the similarities or dissimilarities between oscillations. A similar analysis for case 4 is presented in the panels of the second column.}
\label{fig:standing}
\end{figure}

 One of the critical aspect is to determine the mode of the observed decayless oscillation. To check this, we compare the phase difference between the oscillations captured via slits that are placed at a different position along the length of a loop. 
 In Fig~\ref{fig:standing} we present two example, case-2 and case-4. In each case, we find no phase difference between oscillating threads, indicating that the observed oscillations are fundamental modes of standing waves.

\end{appendix}
\end{document}